\documentclass{article}

\usepackage{PRIMEarxiv}

\usepackage[utf8]{inputenc} 
\usepackage[T1]{fontenc}    
\usepackage{hyperref}       
\usepackage{url}            
\usepackage{booktabs}       
\usepackage{amsfonts}       
\usepackage{nicefrac}       
\usepackage{microtype}      
\usepackage{lipsum}
\usepackage[numbers]{natbib}
\usepackage{fancyhdr}       
\usepackage{graphicx}       
\usepackage{siunitx}
\usepackage{todonotes}
\usepackage{caption}
\usepackage[english]{babel}
\usepackage{epsfig}
\usepackage{amsmath}
\usepackage{float}
\graphicspath{{media/}}     
\usepackage{tikz,xcolor,hyperref}
\definecolor{lime}{HTML}{A6CE39}
\DeclareRobustCommand{\orcidicon}{%
    \begin{tikzpicture}
    \draw[lime, fill=lime] (0,0) 
    circle [radius=0.16] 
    node[white] {{\fontfamily{qag}\selectfont \tiny ID}};    \draw[white, fill=white] (-0.0625,0.095) 
    circle [radius=0.007];    \end{tikzpicture}
    \hspace{-2mm}}
\foreach \x in {A, ..., Z}{%
    \expandafter\xdef\csname orcid\x\endcsname{\noexpand\href{https://orcid.org/\csname orcidauthor\x\endcsname}{\noexpand\orcidicon}}
    }
\newcommand{\eq}[1]{(\ref{#1})}
\newcommand{\fig}[1]{Fig. \ref{#1}}

\newcommand{\be}{\begin{equation}}
\newcommand{\ee}{\end{equation}}

\title{Spontaneous synchronization of two bistable pyridine-furan nanosprings connected by an oligomeric bridge.
}

\author{Anastasia A. Markina $^{1,2,a}$\orcidA, Maria A. Frolkina$^{1,2}$\orcidD, Alexander D. Muratov$^{1,2}$\orcidE, Vladislav S. Petrovskii $^{1,2}$\orcidF,\\ \textbf{Alexander F. Valov$^{1,2}$\orcidC, and Vladik A. Avetisov $^{1,2,b}$\orcidB{}}\\
$^{1}$ \quad N. N. Semenov Federal Research Center of Chemical Physics, Russian Academy of Sciences, Kosygina 4,\\
119991 Moscow, Russia;\\ 
$^{2}$ \quad Design Center for Molecular Machines, Moscow, Russia;\\
  \texttt{$^{a}$ markinanasty@mail.ru} \\
  \texttt{$^{b}$ avetisov@chph.ras.ru} \\
}

\begin{document}
\maketitle

\begin{abstract}
The intensive development of nanodevices acting as two-state systems has motivated the search for nanoscale molecular structures whose long-term conformational dynamics are similar to the dynamics of bistable mechanical systems such as Euler arches and Duffing oscillators. Collective synchrony in bistable dynamics of molecular-sized systems has attracted immense attention as a potential pathway to amplify the output signals of molecular nanodevices. Recently, pyridin-furan oligomers of helical shape that are a few nanometers in size and exhibit bistable dynamics similar to a Duffing oscillator have been identified through molecular dynamics simulations. In this article, we present the case of dynamical synchronization of these bistable systems. We show that two pyridine-furan springs connected by a rigid oligomeric bridge spontaneously synchronize vibrations and stochastic resonance enhances the synchronization effect.
\end{abstract}


\section{Introduction}
Spontaneous synchronization is a phenomenon in which a number of coupled dynamical systems self-organize to behave consistently over time. Synchronization phenomena are observed in various physical, chemical, and biological systems over a wide range of spatial and temporal scales (see, for example, \citeauthor{Neiman1999SynchronizationOT, VanWiggeren1998CommunicationWC, jung_collective_1992,gandhimathi_vibrational_2006,kenfack_stochastic_2010} or \citeauthor{muralidharan_broadband_2022}\cite{Neiman1999SynchronizationOT, VanWiggeren1998CommunicationWC,jung_collective_1992,gandhimathi_vibrational_2006,kenfack_stochastic_2010,muralidharan_broadband_2022}). The concept of synchronization is especially important in the design of nanoscale devices since there is often a need to amplify the weak output signals from individual molecular-size functional units while maintaining the high sensitivity of the device to weak stimuli.

A canonical example of a spontaneously synchronized system is a set of  mechanical or electric oscillators that become coupled with each other via interacting forces.\cite{araki_self-entrainment_1975,strogatz_kuramoto_2000,dorfler_synchronization_2014,tang_synchronization_2014,joshi_synchronization_2016}. Another canonical example is the spontaneous synchronization of spontaneous vibrations (SV) of coupled bistable systems, i.e. synchronization of the noise-activated random transitions between the states of individual bistable systems. \cite{neiman_synchronizationlike_1994}. Unlike an oscillatory system that has eigenfrequency, SV of a bistable system is characterized by a wide frequency distribution\cite{benzi_mechanism_1981,gammaitoni_stochastic_1998,wellens_stochastic_2004}. Nevertheless, SV of coupled bistable systems can also exhibit spontaneous synchronization  \cite{neiman_synchronizationlike_1994,shiino_dynamical_1987} despite the fact that the transitions between the states of the bistable systems remain random. 

In our recent papers \citet{avetisov2019oligomeric,markina2020detection,avetisov2021short}) we have shown that nano-sized oligomeric structures stabilized by short-range low-energy interactions, such as weak hydrogen bonds, hydrophilic-hydrophobic interactions, and  \( \pi\) stacking, could exhibit bistability and SV caused by fluctuations coming from the ambient heat bath. Using molecular dynamic simulations, we demonstrated that short pyridine-furan oligomers a few nanometers in size (named oligo-PF springs), which helical shape was stabilized in water by the \( \pi-\pi\) interactions of aromatic groups, exhibited bistability, SV, and the stochastic resonance (SR) effects (see \citet{avetisov2021short}).  All these effects are characteristic of the classical example of the springs with nonlinear elasticity known as Duffing oscillators. As described in (see \citet{avetisov2021short}), these effects appeared in oligo-PF springs due to the contribution of \( \pi-\pi\) interactions into the spring stretching energy. 

Bistable oligo-PF springs, being two-state dynamic systems, are attractive for applications because the thermal fluctations present at room temperature are sufficient for them to work. However, experimenting with such a small spring and its implementation as an operational unit is challenging since it might require detection extremly weak input signals and sophisticated single-molecule manipulations. It seems desirable to amplify a weak response of an individual bistable nanospring but retain those inviting characteristics that are due to the nanosize of the springs, such as bistability and spontaneous vibrations activated by normal thermal noise from the surrounding thermal bath. This is a non-trivial task because a simple lengthening of an PF-nanospring up to several turns makes it's vibrational dynamics multimode due to extra degrees of freedom of movements of the turn relative each other. As a result, the spring bistability itself becomes ambiguous. In addition, an increase in the size of the nanospring, even maintaning bistability, would lead to a rise in the activation barrier for SV, and the finest sensitivity of the nanospring to weak perturbations could be lost. Therefore, if one wants to scale the finest characteristics of bistability of individual nanosprings, then coupling and synchronization of the nanosprings seems like a suitable solution. The question is what kind of coupling can lead to the synchronization of SV of oligo-PF springs.

In this paper, we present the simulations of long-term molecular dynamics of two bistable oligo-PF springs in water, which are coupled to each other by an oligo-pyrrole bridge. We show that such coupling leads to significant synchronization of the SV of the nanosprings. We also show that the synchronization effect is enhanced in the SR regime of the PF-spring vibrations. The paper is organized as follows. In Section 1, we introduce the theory of spontaneous synchronization of bistable systems as a footing for computer simulation studies of the dynamics of coupled pyridine-furan nanosprings and provide details of the computer model and simulations. In Section 2, we present the computer simulation data and discuss the synchronization effects in the regimes of SV and SR. The paper ends with the discussions and conclusions sections, Sections 3 and 4.
\section{Materials and Methods}

\subsection{Theoretical modelling: Langevin approximation}\label{Sec:MatMet1}

Here we introduce the theoretical underpinning of the bistable system's synchronization to align the computer simulation studies presented below with the concept of synchronization. First, let us consider the overdamped regime of two coupled identical Duffing oscillators firstly discussed by Neiman in \cite{neiman_synchronizationlike_1994}. The two-dimensional dynamics of the system in dimensionless units is described by Langevin equations:

\be
\begin{cases}
\frac{dx}{dt}=-\frac{dU(x,t)}{dx}+\epsilon\xi_1(t) +G_1(x,y) \\
\frac{dy}{dt}=-\frac{dU(y,t)}{dy}+\epsilon\xi_2(t) +G_2(x,y),
\end{cases} 
\label{sde}
\ee

where $x$ and $y$ are the degrees of freedom of the first and second Duffing oscillator, respectively, $\xi_i(t)$ is a delta-correlated white noise, $\epsilon$ is a noise amplitude, \(U(x)=-\frac{\alpha}{2}x^2+\frac{1}{4}x^4 +E_0 x \cos(\omega t)
\) is a periodically modulated double well potential. The parameter $\alpha$  determines the positions of the potential minima, $x_{min}=\pm\sqrt{\alpha}$, and the bistability barrier, $\Delta U=\alpha^2/4$, separating the minima. The coupling forces $G_i(x,y)$ are introduced in the form:
\be
\begin{cases}
G_1(x,y)=b(y-x) \\
G_2(x,y)=b(x-y)
\end{cases} 
\label{CF}
\ee
where $b>0$ is the coupling constant. It is easy to see that the coupling tries to synchronize the dynamics of bistable systems by minimization of the difference between $x$ and $y$. 

In what follows, we will use numerical simulations of the equations (1) using, e.g.,  Heun’s method, to represent the synchronization effect. Specifically, we choose dimensionless parameters $\alpha=10$, $\epsilon^2=18$ for which the Kramers rate $r_K\propto\exp\left(-\frac{\alpha^2}{2\epsilon^2}\right)$ is close to those obtained in molecular simulations, and we varied coupling constants $b$ from $b=0-0.75$. Given potential parameters, the critical value of the external field, at which the energy injected by the field  in one period is comparable to the barrier value, is $E_{cr}=\Delta U/x_{min}\approx 7.9$.

In the absence of coupling and periodic modulations ($b=0$ and $E=0$), the random nose activates jump-like transitions over the bistability barrier with with the mean life time of the states corresponding to Kramers rate.

\begin{figure}[ht]
\centering
\includegraphics[width=.95\textwidth]{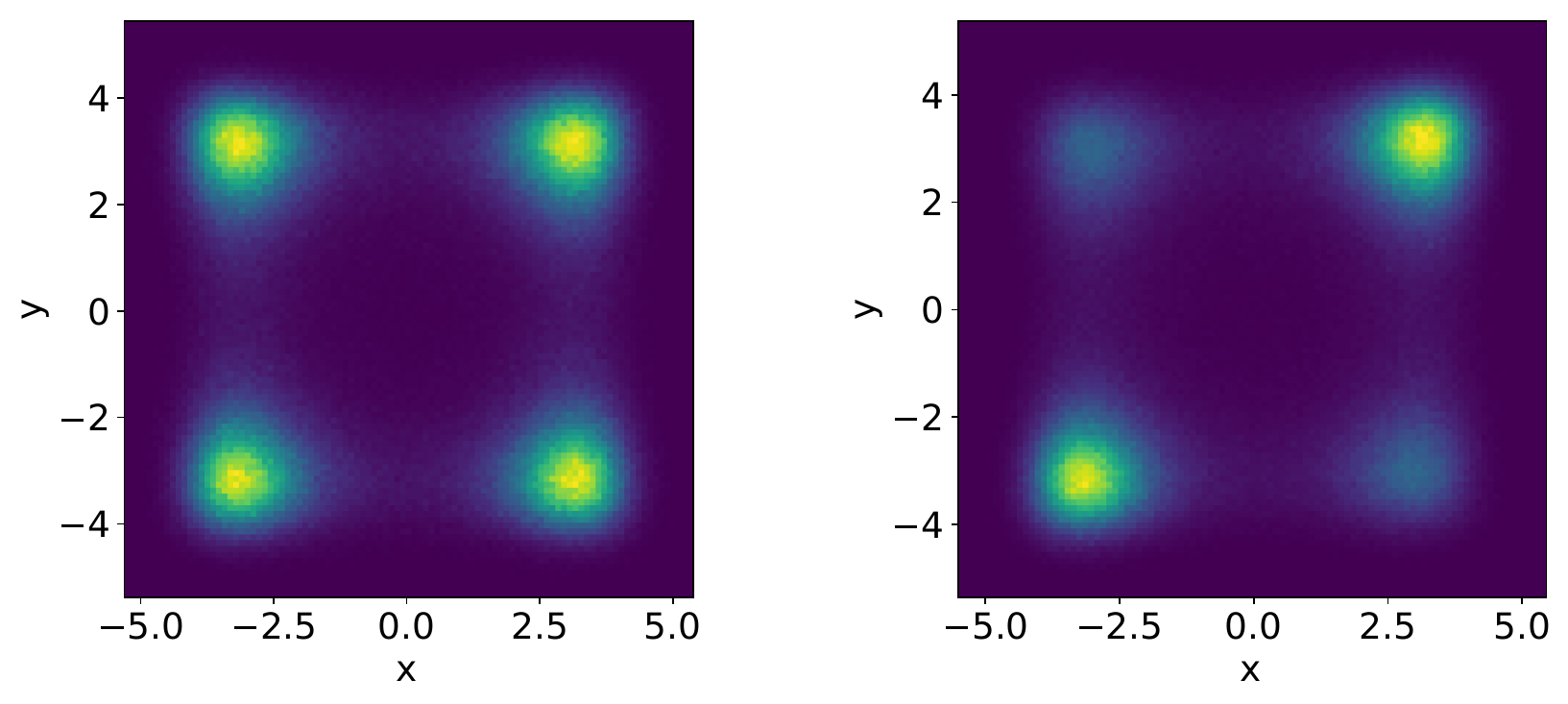}
\caption{ The 2D  stationary density $P(x,y)$ to find the first and the second bistable oscillators in the position $x$ and $y$, respectively,  at different values of coupling  (lighter colors correspond to higher probability density and dark violet means zero probability density): left panel - SV of independent bistable systems ($b=0$ and $E=0$); right panel - SV of coupled bistable system ($b=0.5$ and $E=0$).}
\label{fig02b}
\end{figure}

In this case, the system of two Duffing oscillators has four attractors shown on the left panel of Fig. 1, which correspond to the independent SV of the Duffing oscillators. The same brightness of the attractors indicates that the system equally visits them. In the case of coupling (\fig{fig02b}, right panel), two attractors are visited more often, indicating that random transitions of two Duffing oscillators are synchronous.

A weak periodic field applied to the system (see \eq{sde}) cyclically tilts the bistable potential of each Duffing oscillator and induces periodic modulation of the Kramers rate of the transitions between the two states. It is well known that if the period of modulation coincides with the double inverse $r_{K}$, the SR occurs, and the random transitions between the two states become highly regular. We refer readers interested in fine-tuning the SR  time-scale matching conditions to \citet{gammaitoni_stochastic_1998} and references therein. 

\begin{figure}[ht]
\centering
\includegraphics[width=.55\textwidth]{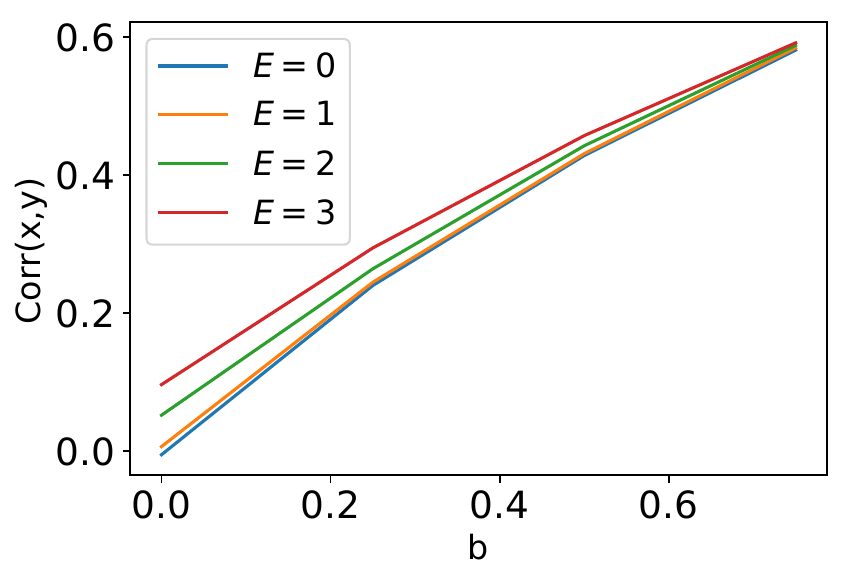}
\caption{The correlations \(corr(x,y)\) of vibrations of two coupled Duffing oscillators  as functions of coupling constant $b$ at different amplitudes  $E$ of the oscillating field modulating the transitions between two states of Duffing oscillator}
\label{fig01corr}
\end{figure}

It is clear that modulation of Duffing oscillators by a periodic field should make an additional contribution to the synchronization effect. However, even in SR mode, the improvement in synchronization depends much more on the coupling than on the modulating field (see \fig{fig01corr}).

\subsection{Pyridine-Furan Springs}
As described in the previous section, the synchronization is achieved by an appropriate coupling of two bistable systems. In our previous paper \cite{avetisov2021short} we have shown that oligo-PF springs a few nanometers in size subjected to critical tension exhibit bistable dynamics characteristic of Duffing oscillators. Accordingly, they were also chosen for this paper to study the synchronization phenomena.

\begin{figure}[H]
\includegraphics[width=13.5 cm]{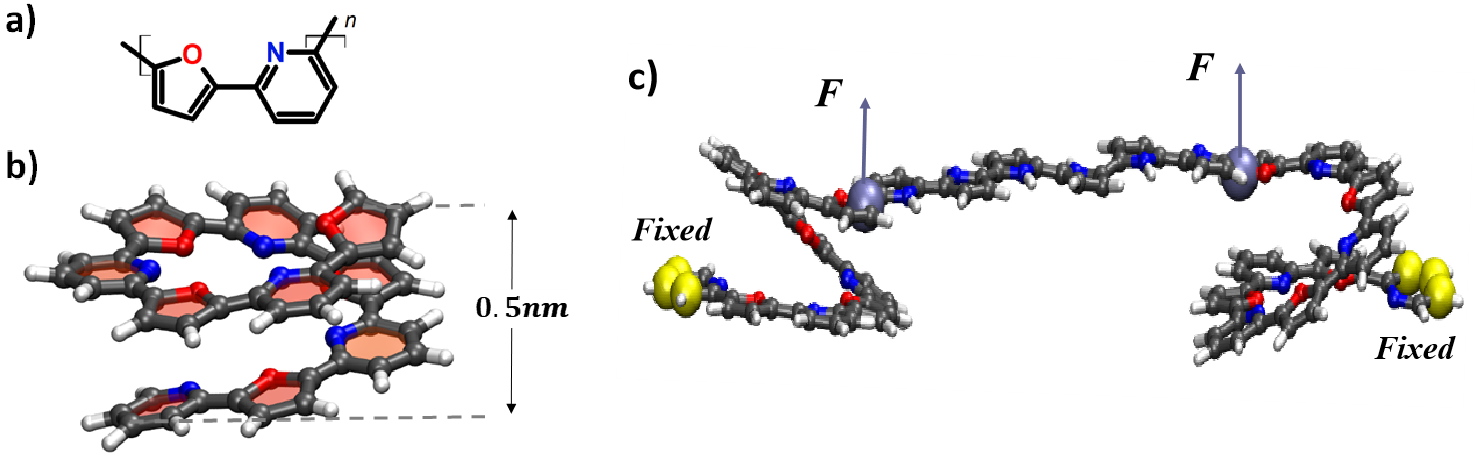}
\caption{($\mathbf{a}$) Chemical structure of a pyridine-furan monomer unit with heterocyclic rings in cis-configuration; ($\mathbf{b}$) a single pyridine-furan spring five monomer units long (oligo-PF-5); ($\mathbf{c}$) Two PF-nanosprings connected with an oligo-pyrrole bar of the length of six monomer units, the tensile forces are shown as vectors \(F\) (applied to the top ends of each spring), the yellow spheres at the bottom (conditionally) end of the spring indicate the fixation of the pyridine ring by rigid harmonic forces.\label{fig1}}
\end{figure}

Pyridine-furan copolymer (Figure ~\ref{fig1}a) is a conductive polymer consisting of 5- and 6-member heterocyclic pyridine and furan rings (respectively)\cite{jones1997extended}. These PF copolymers in the cis-configuration tend to form a helix-like shape, which is stabilized by the \( \pi-\pi\) interactions of aromatic groups located on the adjacent turns of the spring \cite{sahu2015}. In particular, the cis-configuration of oligo-PF of 5 monomer units, oligo-PF-5, was selected for study in this paper (see Figure 1a-b). Since the coupling between the two oligo-PF springs was supposed to be like a rigid oligomeric bar, the various chains for the bar were tested, and an optimal option was defined as an oligo-pyrrole chain of six monomer units in length (see Figure ~\ref{fig1}), which provided a smooth but rigid connection between both springs. 

\unskip

\subsection{Simulation Details}
Molecular dynamics is the best simulation method available to model effects like spontaneous synchronization where the composite behavior of the system depends on the physical motion of molecules. The PF-nanospring system was modeled in a fully atomistic representation in the canonical ensemble (number of particles/volume/temperature [NVT]). The springs were surrounded by enviornmental water, and the total simulation box size was \(7.0 * 7.0 * 7.0\) \(nm^3)\). Simulations were conducted with a time step of 2 fs using the Gromacs 2019 package \cite{abr2015}, OPLS-AA \cite{kam2001} force field parameters for the oligomer, and the SPC/E model \cite{ber1987} for water (for more details, see Parameters for \emph{Molecular Dynamics Simulation section} of \emph{Supplementary Materials}). Following characterization of the equilibrated state of PF-springs at $\SI{280}{\kelvin}$ \cite{sahu2015}, this temperature was set by the velocity-rescale thermostat \cite{bussi2007canonical}. Each dynamic trajectory was 300–350 ns long and was repeated three times to obtain better statistics; therefore the effective length of the trajectories was about 1 \(\mu s\) for each sample. 

At first, two oligo-PF-5 springs with fixed bottom (conditionally) ends and movable upper ends connected by an oligomeric bar were equilibrated at $\SI{280}{\kelvin}$. For convenience and in accordance with the designations introduced in Subsection \ref{Sec:MatMet1} we call (conditionally) left spring $X$ and right spring $Y$. Then, springs were subjected to an external tensile forces \emph{F} (in Figure ~\ref{fig1}c): at that, the bottom ends of the springs were kept fixed, and the upper ends were pulled by forces applied normally to the connecting bar. The end-to-end distance of each of the oligo-PF-5 springs, $R_{e}^{X}$ and $R_{e}^{Y}$, (see Figure~\ref{fig1}b) was chosen as an order parameter describing the long-term dynamics of the spring.  Vibrations of coupled oligo-PF-5 springs occur between two well-reproduced states of each spring with end-to-end distances equal to \(R_e \approx 1.0 nm\) and \(R_e \approx 1.4 nm\). These states are referred to as the squeezed and the stress-strain states, respectively\cite{avetisov2021short}.
 
The mutual movements of the springs were plotted into 2D phase diagrams, which were the long time statistics of the 2D-probability density for simulteneously visiting  states $R_{e}^{X}$ and $R_{e}^{Y}$, respectively.  In Figure~\ref{fig2} and ~\ref{fig3}, brighter colors correspond to higher probability density and dark violet means zero probability density. The statistics of the two states map were extracted directly from the \(R_e(t)\) series. The synchronization coefficient was defined as the height of the main peak of the normalized cross-correlation function of time trajectories of $R_{e}^{X}$ and $R_{e}^{Y}$.

\section{Results}
\subsection{Synchronization of spontaneous vibrations of PF-5 springs}
Next, the dynamics of the oligo-PF-5 springs connected by an oligomeric bridge were examined as they were subjected to pulling forces of different values. Under weak tension, the springs were stretched slightly in accordance with linear elasticity. However, as soon as the pulling force reached a critical value \(F= 220 pN\), the coupled oligo-PF-$5$ springs became bistable and started to vibrate spontaneously similar to SV of a single oligo-PF-$5$ spring described in detail in our previous paper\cite{avetisov2021short}. Above this critical load, the springs spontaneously vibrate between the squeezed and stress–strain states. SV are observed in a wide region of pulling forces from \(220 pN\) to \(300 pN\). In the regime of SV, the end-to-end distance fluctuations of each individual PF-spring have a mean amplitude of 0.2 nm, so the stress–strain states of both springs are clearly distinguished from the squeezed states. 

\begin{figure}[H]
\includegraphics[width=13 cm]{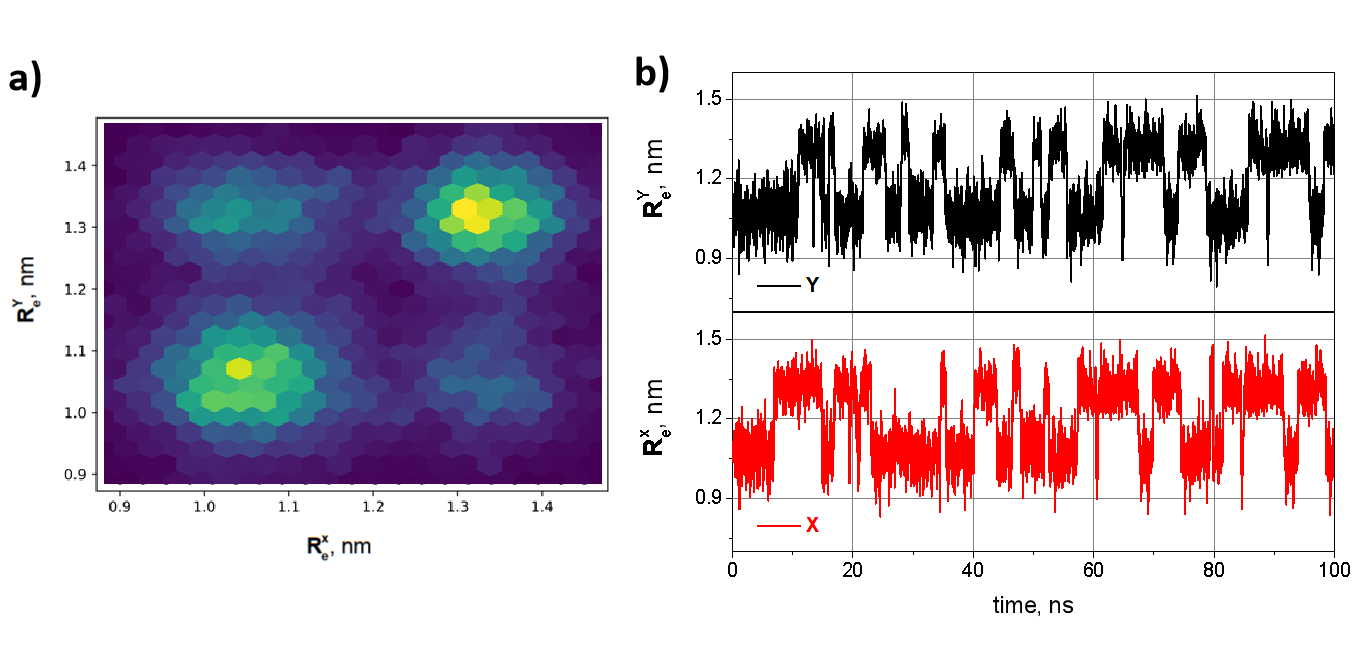}
\caption{SV of two coupled oligo-PF-5 springs at $F=250$ pN: (\textbf{a}) 2D stationary probability density for simultaneous location of the left and right springs at the positions $R_{e}^{X}$ and $R_{e}^{Y}$, respectively; (\textbf{b}) corresponding trajectories of the end-to-end distances $R_{e}^{X}$ and $R_{e}^{Y}$. 
\label{fig2}}
\end{figure}

Panel $\mathbf{a}$ of Figure~\ref{fig2} shows 2D-phase diagram of two copled PF-springs, i.e., 2D stationary probability density for simultaneous location of the ends of  left and right springs at the positions $R_{e}^{X}$ and $R_{e}^{Y}$, respectively. Figure~\ref{fig2}$\mathbf{b}$ shows a shortened part of typical SV-trajectories (for the full trajectory see the Supplementary Materials) of the coupled oligo-PF-5 springs in the symmetric bistability region (\(F = 250\)) pN. In this regime, SV are very pronounced, because neither the squeezed nor stress–strain states dominate, so the mean lifetimes of the two states for both springs were approximately the same, and equal to \( \tau = 6.14\) ns. The bistability barrier of the oligo-PF-5 springs is circa 10  \(k_BT\), as was shown in our previous paper\cite{avetisov2021short}. Therefore, twith the except of value of the critical pulling force, all other characteristics of bistability of individual PF-springs in coupled system. e.g. the amplitude of vibrations, mean lifetimes of the squeezed state and the stress-stain  state, the intensity of thermal noise activating the  spontaneous vibrations, etc.,  are close to the same characteristics of a freestanding PF-5 spring.

\begin{figure}[H]
\includegraphics[width=13.5 cm]{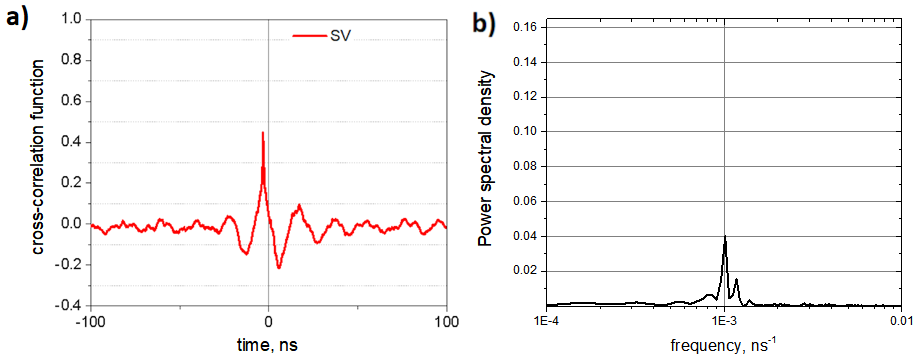}
\caption{Correlation between SV of the PF-nanosprings coupled by an oligomeric bar: ($\mathbf{a}$)  cross-correlation function for the SV-trajectories of the left and right springs; ($\mathbf{b}$) power spectrum of the cross-correlation function.}
\label{fig3-1}
\end{figure}

Figure ~\ref{fig3-1} a shows cross-correlation function between SV-trajectories of the left and right springs coupled by an oligomeric bar. The power spectrum of this function is given on Figure 5b. The cross-correlation function clearly demonstrates that spontaneous vibrations of coupled oligo-PF-5 nanosprings are significantly synchronized with the correlation coefficient $k \approx 0.5$. The power spectrum of the autocorrelation function indicates that the spectra of spontaneous vibrations of both nanosprings do not have natural frequencies, with the exception of a weak specific response near the so-called SR-frequency, the reciprocal of which is equal to twice the lifetime of the state when the system experiences spontaneous oscillations \cite{benzi_mechanism_1981,gammaitoni_stochastic_1998,wellens_stochastic_2004}.

\subsection{Synchronization of the spring vibrations in the stochastic resonance regime.}

\begin{figure}[H]
\includegraphics[width=13 cm]{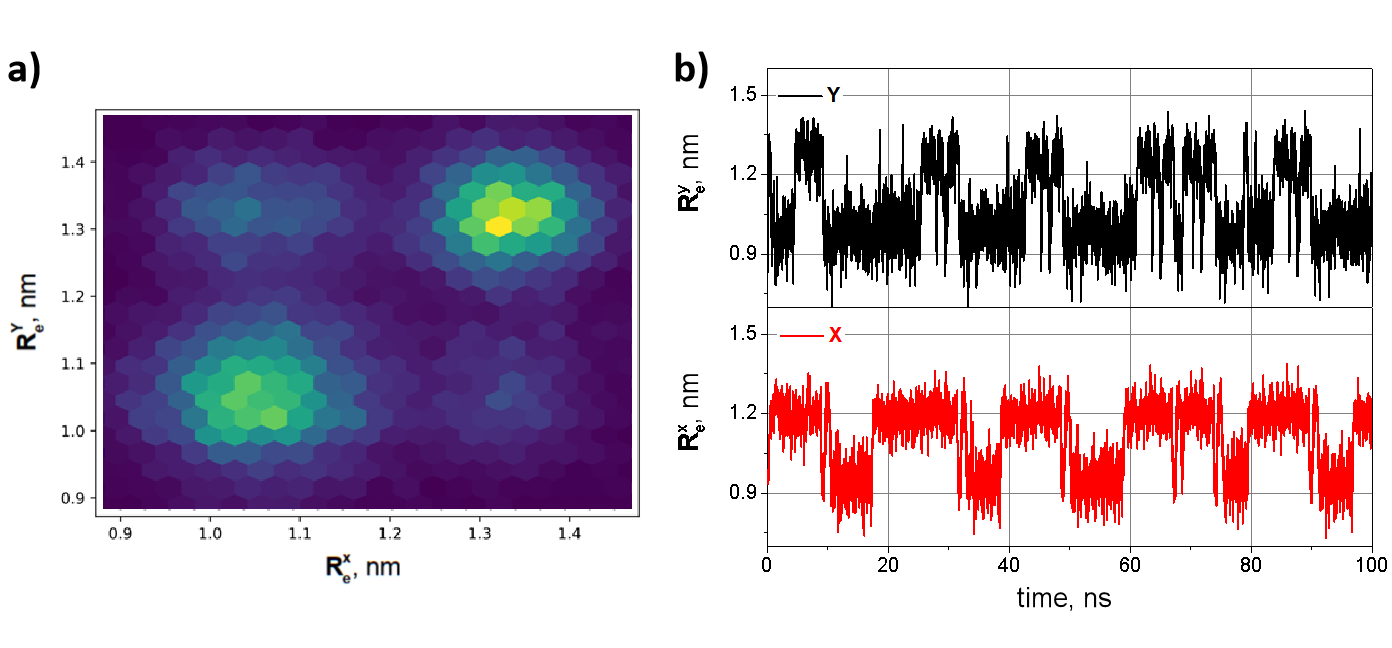}
\caption{Vibrations of the oligo-PF-5 springs coupled by an oligomeric bar in the SR-regime: (\textbf{a}) 2D stationary probability density for simultaneous location of the left and right springs at the positions $R_{e}^{X}$ and $R_{e}^{Y}$, respectively; (\textbf{b}) typical SR-trajectories $R_{e}^{X}$ and $R_{e}^{Y}$ of individual PF-5 springs.
\label{fig3}}
\end{figure}

Next, we examined the SR mode of the coupled oligo-PF-5 springs by applying an additional oscillating forces that weakly modulated the upper ends of the coupled springs. The oscillating electrical field \(E = E_0 \cos(2\pi \nu t)\), \(E_0\) = 0.1 V$/$nm, \( \nu = 1/12.5\) GHz acting on unit charges present at the upper ends of the springs was used to model SR, while the compensating charges were at the bottom (fixed) ends of the springs (for more details, see Parameters of the periodic signal section of Supplementary Materials). Note, that the frequency \( \nu = 1/12.5\) GHz is the exactly the SR frequency equal to the half of the Kramers frequency of the transitions over the bistability barriers of individual PF-springs. Typical vibrations of coupled oligo-PF-5 springs in the SR mode are shown on Figure~\ref{fig3}.

\begin{figure}[H]
\includegraphics[width=13.5 cm]{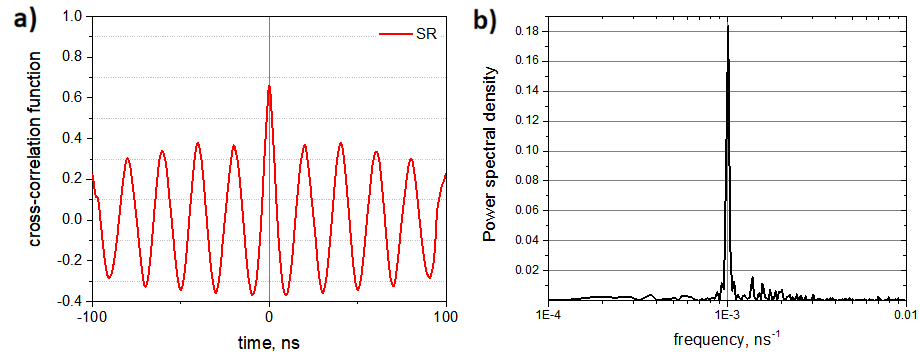}
\caption{Correlation between vibrations of the PF-springs coupled by an oligomeric bar in the SR regime: ($\mathbf{a}$) cross-correlation function of the vibrations the left and right springs; ($\mathbf{b}$) Power spectrum of the cross-correlation function.}
\label{fig3-2}
\end{figure}

The cross-correlation function between the SR-trajectories of the left and right springs coupled by an oligomeric bar is shown in Figure ~\ref{fig3-2} a. The power spectrum of this function is shown on Figure ~\ref{fig3-2} b. The cross-correlation function clearly demonstrates that SR enhance synchronization effect: the correlation coefficient is raised up to $k \approx 0.7$. In its turn, the power spectrum of the correlation function indicates that close to the half of the Kramers frequency of the transitions over the bistability barriers the vibrations of the PF-springs have well-defined SR signal. Thus, the synchronization effect can be improved by applying a weak oscillating force with SR-frequency.

\unskip

\section{Discussion}
Based on the results, the main finding is that coupled oligo-PF-$5$ springs exhibit synchronized dynamics in SV and SR regimes. Similar to a single spring, two coupled springs under the critical stretching show SV. The value of the critical tensile in the case of coupled springs is slightly lower than for a single oligo-PF-$5$ spring (as previously reported\cite{avetisov2021short}). The difference in critical tensiles and dynamic behavior between single and coupled oligo-PF-$5$ springs arises since coupled springs are bridged by a connectivity bar, which causes additional tensiles as compared to single springs. This connectivity bar could form an additional \( \pi\) bond with the springs and thus affect the bistability area and the SV scale. This means that the collective dynamics of any hybrid system comprised of a complex set of coupled springs should be checked independently since specific couplings could introduce new particular tensions and additional intramolecular interactions.

When spontaneous vibrations occur, the presence of a connectivity bar allows the non-correlated jumping behavior to transform into correlated transitions. This would represent a transition either from a squeezed state to a stress-strain state and vice-versa. The correlation coefficient associated with a such transition is circa $0.5$ (i.e. partially correlated regime). Notably, the correlation can be improved in the regime of SR by an introduction  of a weak external force. The external force causes a well-defined frequency and amplitude to appear that characterizes the system dynamics. In this regime, the correlation coefficient grows up to $0.7$ (i.e. high correlation regime). 

These findings are consistent with the mathematical model described in Section \ref{Sec:MatMet1} in terms of synchronization upon introduction of the bond between the bistable systems. However, Figure~\ref{fig01corr} shows that at the SR regime the correlation coefficient between the bistable systems has a very weak dependence on the modulation amplitude \(E\). In other words, there is no significant difference in correlation coefficient in regimes of SV (when \(E=0\)) and SR (\(E>0\)). Also, there are no shifts between the trajectories in the SV regime. On the other hand, in the simulated system the correlation coefficient differs significantly between the SV and SR regimes and at the same time the trajectories of SV for left and right springs were observed to be time-shifted by 3 ns (the peak of the cross-correlation function for $R_{e}^{X}$ and $R_{e}^{Y}$ isn't located at zero). The shifted maximum of the cross-correlation function indicates a systematic time delay in the dynamics of the springs. The finite rigidity of the connectivity bar and the system's geometrical freedom are the reasons why one spring doesn't feel the displacement of another spring immediately. As a result, one of the springs becomes a driver for the synchronous behavior of both. Importantly, no shift is observed in the SR regime since the movement of each spring was as well consistent to the oscillations of the weak external electrical field. Such difference between the theory and simulations can be explained due to the mathematical model of the bond, which is in theory infinitely rigid and immediately transfers the displacements between the bistable systems.

Based on the results presented above, the following best practices can be recommended for designing a particular connecting bar considering the key variables of rigidity and length. In particular, to build a working bistable and controllable structure, it is important to first select a connectivity bar of high rigidity (i.e. high persistent length). This can be a broad class of \( \pi\)-conjugated structures which includes pyrrole, as used in this study. The more rigid connection is introduced, the stroger coupling of bistable systems is achieved and thus the synchronization effect increases.

The second crucial design factor is the connectivity bar's length. A small bar obstructs the mutual motion of the coupled springs due to sterical hindrance. This hindrance occurs due to the complex nature of spring motion. The end of spring is moving not only up, but also a bit sideways (motion reminding a helical path). These tiny sideways motions are oppositely directed for left and right springs, thus causing slight rotation in the plane of the connectivity bar around its center. If the connectivity bar is short, the rotation angle is limited, and therefore sideways movements are limited. As a consequence vertical movements are locked as well.   

A connectivity bar of longer length will avoid sterical obstruction. However, an increase of the length of the connectivity bar naturally causes a decrease in its rigidity and causes a delay between the displacements of left and right oligo-PF springs. A large connectivity bar will work as a big lever arm and force springs to move further from their equilibrium, causing an artificially large amplitude. In other words, interplay between the elastic properties of the oligo-PF springs and the torsional deformation of the crossbar are defining the overall system dynamics. Taking into account both these factors, choosing a semirigid bar of circa 2 nm in size would be preferred - one that is rigid enough to provide a bond but does not influence the motion of the springs. Future conjugated structures that have high persistence lengths could be considered as candidates for a best-performing coupling.


\section{Conclusions}
In this paper, we have presented a study of the mechanic-like behavior of a novel nanospring system composed of two bistable oligomeric pyridin-furan springs coupled by an oligomeric pyrrole bridge. Through molecular dynamics simulations in all-atom representation, we have investigated the long-term conformational dynamics of the system and found that the sponteneus vibrations of coupled springs become synchronized. 

The system exhibits synchronized vibrations under certain conditions, which can be controlled, for example, by changing the tensile forces, the length and stiffness of the oligomeric bridge, and the elasticity of the springs. The simulations of two PF-nanospring coupled by an pyrrole bridge show that SV of the springs are synchronized with the correlation coefficient $k\approx0.5$. In the case of spontaneous synchronization of oligo-PF-5 springs in the SR mode, the correlation coefficient reaches a relatively high value of $k\approx0.7$. In summary, our study provides new insights into the bistable nanomechanics of coupled PF-nanosprings and highlights the fact that the synchronization can be constructively achieved in relatively simple ways.

\section*{Funding}
The authors acknowledge the financial support of the Design Center for Molecular Machines, Moscow, Russia, under the Business Contract $28$/$11$-X$\Phi$-$28.11.2022$ with N. N. Semenov Federal Research Center for Chemical Physics, Russian Academy of Sciences.
\section*{Acknowledgments}
We thank Alexey Astakhov, Vladimir Bochenkov and Dmitry Pergushov for helpful discussions.


\end{document}



\section{\textbf{Simulation protocol}}
\begin{figure}[H]
\includegraphics[width=10.5 cm]{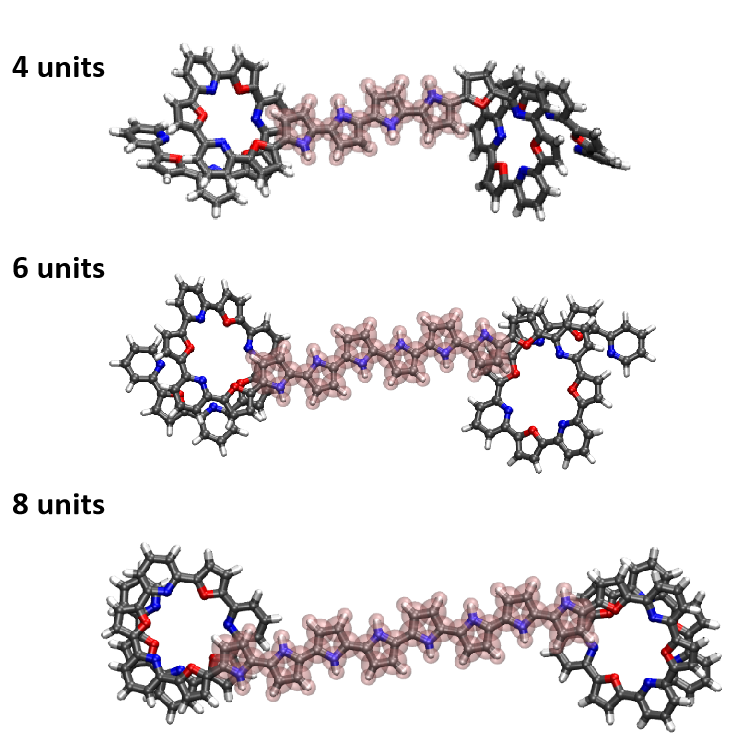}
\caption{Two oligo-PF-5 springs connected with oligo-pyrrole bars of different lengths: 4, 6 and 8 units.\label{figS1}}
\end{figure}

Morphology simulations were performed using the GROMACS simulation package. Details of the interaction parameters and experiment procedure are described elsewhere [17-19]. All calculations were performed in the NVT ensemble, in the simulation box of circa 52000 water molecules and size of 7nm x 7nm x 7nm using the canonical velocity-rescaling thermostat, as implemented in the GROMACS simulation package. The simulation was started from a random initial configuration and run to reach an equilibrated morphology. To study how the oligomers respond to a power load, the system simulation was continued for an additional 800 ns. The model error was estimated using the full width at \(50\%\) of the distribution curve maximum. To show that the results are independent of the specific bar length, different bar lengths were tested (see Figure~\ref{figS1}). 

SV are observed for all three cases in a quite wide region (from 140pN to 220pN). The short connectivity bar (4 units) synchronizes the springs. However, the two states are not very well defined and overlapped on the histogram, due to steric blockings from the bar side (see Figure~\label{figS2}). In order to stretch springs are going up and at the same time rotate according to directions shown by red arrows. With a longer connectivity bar of length 8 relative movements of two springs are not blocked. In this case, two states are defined better and do not overlap significantly on the histogram. The median length bar gives a better balance between the synchronization and splitting of two states.

\begin{figure}[H]
\includegraphics[width=13.5 cm]{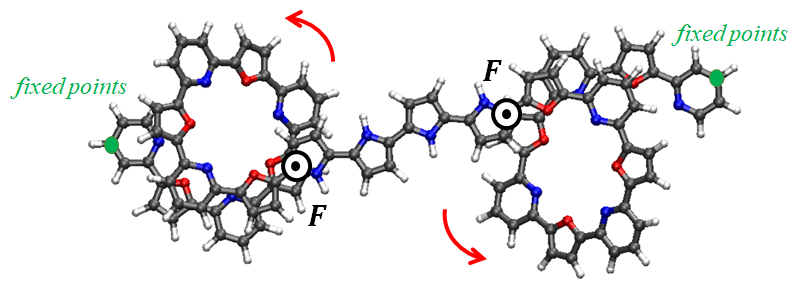}
\caption{Two oligo-PF-5 springs connected with a short oligo-pyrrole bar of 4 units under the external load. Red arrows show the relative rotation of each spring. \label{figS2}}
\end{figure}   

\section{\textbf{Spontaneous vibrations and Stochastic Resonance}}

\begin{figure}[H]
\includegraphics[width=13.5 cm]{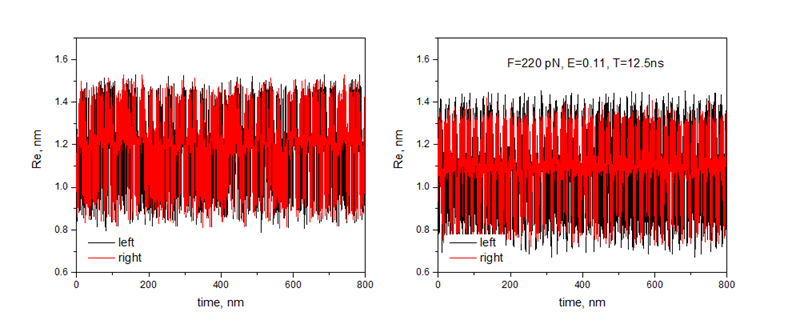}
\caption{Example of trajectories of the end to end distance of two oligo-PF-5 springs connected with a long oligo-pyrrole bar of 6 units. (\textbf{left}) SV of the springs at F=220 pN; (\textbf{right}) SR of the springs induced by an oscillating field $E = E_{0} \cos(2\pi\nu t) =E_{0} \cos(2\pi t/T)$.\label{figS3}}
\end{figure}   

\begin{figure}[H]
\includegraphics[width=13.5 cm]{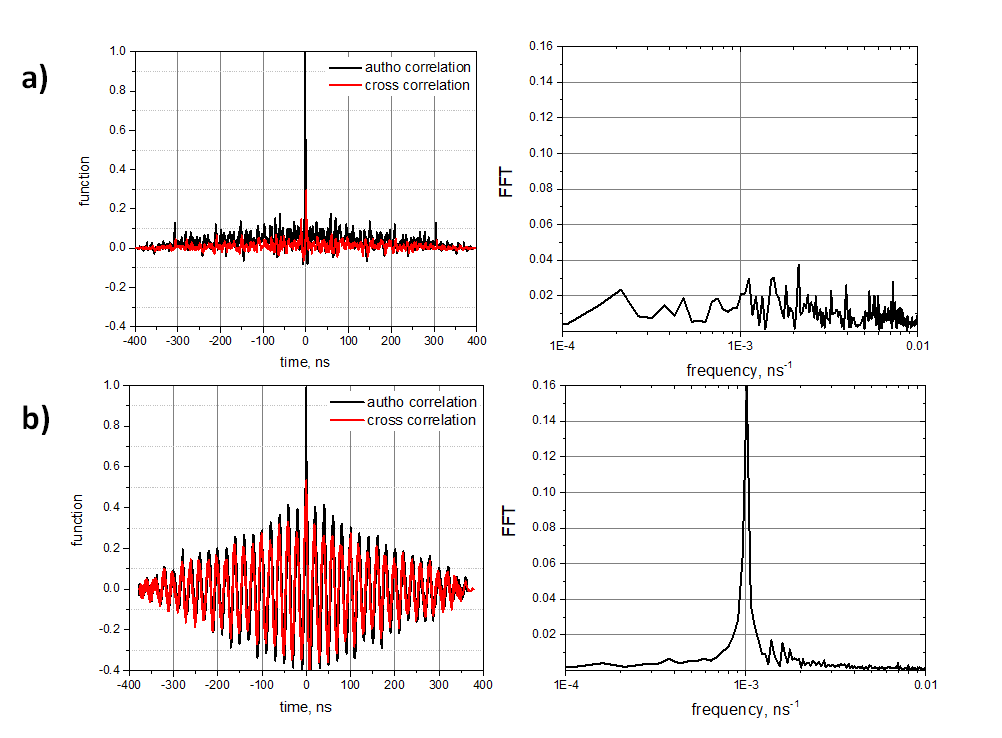}
\caption{(\textbf{a)}) SV of the springs at $F \approx 220$ pN: correlation and autocorrelation functions of two springs (on the left) and Fourier spectrum of the autocorrelation function of the right spring (on the right). (\textbf{b)}) System with applied external field (SR): correlation and autocorrelation functions of two springs (on the left) and Fourier spectrum of the autocorrelation function of the right spring (on the right).\label{figS4}}
\end{figure}

\end{adjustwidth}